\documentclass{article}
\usepackage{spconf,amsmath,graphicx}
\usepackage{enumitem}
\setlist{nosep, leftmargin=14pt}
\usepackage{mwe} % to get dummy images
\usepackage[
 colorlinks=false
]{hyperref}
\usepackage{fancyhdr}

\fancypagestyle{specialfooter}{%
  \fancyhf{}
  
  \fancyfoot[C]{\textbf{$\copyright$ 2022 IEEE. Personal use of this material is permitted. Permission from IEEE must be obtained for all other uses, in any current or future media, including reprinting/republishing this material for advertising or promotional purposes, creating new collective works, for resale or redistribution to servers or lists, or reuse of any copyrighted component of this work in other works.}}
}

\title{Convolutional Analysis Operator Learning by
End-To-End Training of Iterative Neural Networks}
\name{Andreas Kofler$^1$, Christian Wald$^2$,  Tobias Schaeffter$^{1,3,4}$, Markus Haltmeier$^5$, Christoph Kolbitsch$^{1,3}$ }
\address{$^1$ Physikalisch-Technische Bundesanstalt, Berlin and Braunschweig, Germany \\ 
$^2$ Department of Radiology, Charité - Universitätsmedizin Berlin, Berlin, Germany \\ 
$^3$ School of Imaging Sciences and Biomedical Engineering, King’s College London, London, UK\\
$^4$ Department of Biomedical Engineering, Technical University of Berlin, Berlin, Germany \\ 
$^5$ Department of Mathematics, University of Innsbruck, Innsbruck, Austria\thanks{Our implementation of the network is available under  
\url{www.github.com/koflera/ConvSparsityNNs}.} }

\usepackage{graphicx}
\usepackage{xcolor}
\usepackage{amssymb,amsmath,amsfonts}
\usepackage{tikz}
\usepackage{multirow}
\usepackage{booktabs}
\usepackage{blindtext}
\usepackage{multicol}
\usepackage{tikzscale}
\usepackage{numprint}
\npdecimalsign{.}
\nprounddigits{3}
\usepackage{parskip}
\usepackage{enumitem}
\usepackage{float}
\usepackage{pict2e}
\usepackage{siunitx}
\usepackage[english]{babel}

 \newcommand{\sd}{\mathbf{s}}                         
  
\newcommand{\XX}{\mathbf{x}}  
                       
\newcommand{\ZZ}{\mathbf{z}}                         
\newcommand{\YY}{\mathbf{y}}                         
                         
\newcommand{\Ad}{\mathbf A}                         
\newcommand{\Wd}{\mathbf W} 
                        
\newcommand{\Id}{\mathbf I}                         
\newcommand{\Cd}{\mathbf C}                         
\newcommand{\Ed}{\mathbf E}                         
\newcommand{\Hd}{\mathbf H}

\newcommand{\Au}{\mathbf{A}_I}

\newcommand{\herm}{{\scriptstyle \boldsymbol{\mathsf{H}}}}
\newcommand{\trans}{{\scriptstyle \boldsymbol{\mathsf{T}}}}

\usepackage[percent]{overpic}
\usepackage{algorithm,algcompatible,amsmath}
\algnewcommand\INPUT{\item[\textbf{Input:}]}%
\algnewcommand\PARAMETER{\item[\textbf{Parameters:}]}%
\algnewcommand\OUTPUT{\item[\textbf{Output:}]}%
\colorlet{lred}{red!80}
\colorlet{cmix}{blue!80!red} 
\colorlet{lgreen}{green!80}
\colorlet{lblue}{blue!80}

\numberwithin{theorem}{section}
\definecolor{amber(sae/ece)}{rgb}{1.0, 0.49, 0.0}
\definecolor{blue(ryb)}{rgb}{0.01, 0.28, 1.0}

\begin{document}
\thispagestyle{specialfooter}
\maketitle
\begin{abstract}
The concept of sparsity has been extensively applied for regularization in image reconstruction. Typically,  sparsifying transforms are either pre-trained on ground-truth images or adaptively trained during the reconstruction. Thereby, learning algorithms are designed to minimize some target function which encodes the desired properties of the transform. However, this procedure ignores the subsequently employed reconstruction algorithm  as well as the physical model which is responsible for the image formation process. Iterative neural networks - which contain the physical model - can  overcome these issues. In this work, we demonstrate how convolutional sparsifying filters can be efficiently learned by end-to-end training of iterative neural networks. We evaluated our approach on a non-Cartesian 2D cardiac cine MRI example and show that the obtained filters are better suitable for the corresponding reconstruction algorithm than the ones obtained by decoupled pre-training. 
\end{abstract}
\section{Introduction}
\label{sec:intro}
Recently, iterative convolutional neural networks (CNNs) have been successfully applied to image reconstruction problems  and seem to define the state-of-the-art across many imaging modalities, see e.g.\ \cite{adler2017solving}, \cite{hammernik2018learning}, \cite{schlemper2017deep}, \cite{hauptmann2018model}. Iterative CNNs resemble iterative reconstruction schemes of finite length in which the regularizer is parametrized by convolutional operations and can be learned in a supervised manner by end-to-end training of the network. Their success seems to be attributable to i) the fact that the physical model is inherently present in the learning process - which has been reported to lower the expected maximum error-bound \cite{maier2019learning} -  and ii) because the regularizers are trained in conjunction with the reconstruction algorithm that is used to reconstruct the images.\\
Regardless of their success, neural networks have also been reported to possibly suffer from instabilities \cite{Antun201907377} and still operate as black-boxes.  This is an issue especially for a field such as medical imaging where the image content directly impacts diagnosis and treatment planning or decisions. In contrast, more classical learning-based regularization approaches typically come with solid mathematical theory, see e.g.\  \cite{chun2017convolutional}, \cite{Chun2020ConvolutionalAO}. However, in these algorithms - in contrast to iterative neural networks - the physical model is not integrated in the learning process and typically,  training refers to minimizing some object function which reflects the desired properties of the regularizer rather than being optimal for the purpose they have to serve in a subsequent reconstruction process. \\
In this work, we combine the best of the two worlds by using iterative neural networks to train a classical data-driven method based on learned sparsifying transforms given as convolutional filters, similar as in \cite{Chun2020ConvolutionalAO}.  In contrast to iterative NNs using many convolutional layers, the role of the learned regularizer is more transparent. Further, unlike in \cite{Chun2020ConvolutionalAO}, where the filters are pre-trained on a set of ground-truth images, in our network the filters are learned to be optimal with respect to the reconstruction algorithm and the number of iterations that the network uses to reconstruct the images and are adapted to the operator of the inverse problem. Our work also differs from \cite{hammernik2018learning} which  stems from the field-of-experts model \cite{roth2009fields} and uses a Landweber iteration. We instead use a splitting approach, and because of the used formulation, the required non-linear activation function is given by the soft-thresholding operator. In addition, the presented approach differs from the work in \cite{chun2020momentum}, where the filters are trained in a greedy fashion (i.e.\ layer-by-layer), and in the application.
\section{Methods}
\label{sec:methods}
\begin{figure}
\includegraphics[width=\linewidth]{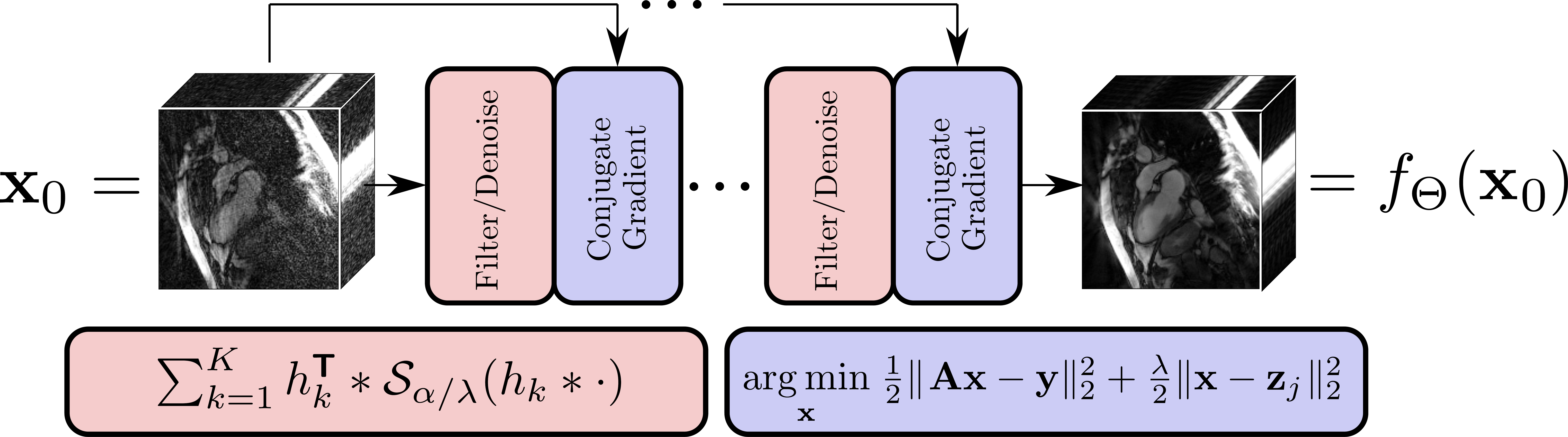}
\caption{Proposed Network structure.  The image is first filtered, soft-thresholded and filtered with the transposed filters. Then, the denoised image is used as regularizing prior in a regularized functional.  
The filters $\{h_k\}_k$ as well as the regularization parameters $\lambda$ and $\alpha$ are obtained by end-to-end training of the entire network.
}\label{fig:network}
\end{figure}

We consider the general type of inverse problem of the form
\begin{equation}\label{eq:inv_problem}
\Ad \XX + \mathbf{e} = \YY,
\end{equation}
where $\Ad$ denotes the forward model, $\XX$ the (unknown) image, $\mathbf{e}$ random noise and $\YY$ the measured data. Problem \eqref{eq:inv_problem} can be ill-posed for different reasons. For example, if $\YY$ has less entries than $\XX$, the problem is underdetermined and there exists an infinite number of solutions. For properly designed overdetermined systems, a solution can be obtained by solving the normal equations, but the stability of the inversion process depends on the condition of the operator $\Ad^\trans \Ad$. In this work, we investigate a regularization method given by the assumption that the image $\XX$ is sparse with respect to convolutional filters. Assuming a \textit{fixed} set of  $K$ sparsifying filters $\{h_k\}_k$, one can formulate the reconstruction problem as a minimization problem
\begin{equation}\label{eq:reco_problem_no_aux}
\underset{\XX}{\min} \frac{1}{2}\|\Ad \XX - \YY\|_2^2 + \alpha \sum_{k=1}^K   \| h_k \ast  \XX\|_1
\end{equation}
over the image $\XX$ with $\alpha>0$. Because $\XX$ is coupled to the operator $\Ad$ as well as to the filters $h_k$ which appear in the $L_1$-norm, directly solving problem \eqref{eq:reco_problem_no_aux}  is challenging. A possible solution strategy is to introduce $K$ auxiliary variables $\sd_k$ to transfer  $h_k \ast \XX$ out of the $L_1$-norm and to relax the equality constraint by including it in a quadratic penalty term, i.e.\\
\begin{equation}\label{eq:reco_problem}
\underset{\XX, \{\mathbf{s}_k\}_k}{\min} \frac{1}{2}\|\Ad \XX - \YY\|_2^2 + \frac{\lambda}{2} \sum_{k=1}^K     \| h_k \ast \XX - \mathbf{s}_k\|_2^2 + \alpha \sum_{k=1}^K   \|\mathbf{s}_k \|_1,
\end{equation}
where $\lambda>0$. 
A possible approach for minimizing \eqref{eq:reco_problem} uses alternating minimization of \eqref{eq:reco_problem} with respect to $\XX$ and $\{\sd_k\}_k$ in an iterative manner \cite{wang2008new}. For fixed $\XX$, problem \eqref{eq:reco_problem} is separable with respect to $k$ and thus, the solution for $\sd_k$ is given by applying the soft-thresholding operator to $h_k \ast \XX$ for all $k$. For fixed $\{\sd_k\}_k$, the minimization with respect to $\XX$ corresponds to solving a linear system $\Hd \XX = \mathbf{b}_j$ with
 \begin{eqnarray}\label{eq:lin_system_alter_min}
\Hd =& \Ad^\herm \Ad +   \lambda \, \sum_{k=1}^K h_k^\trans \ast h_k \\
\mathbf{b}_j =& \Ad^\herm \YY + \lambda \, \sum_{k=1}^K h_k^\trans \ast \sd_k,
\end{eqnarray}  
where we see that the operator $\Hd$ depends on the filters. Since we aim to train the set of filters $\{h_k\}_k$ by training an iterative network in an end-to-end fashion, this alternating-minimization scheme can be compuationally demanding for realistic large-scale applications, e.g.\ for the later discussed dynamic cardiac MRI problem. Therefore, motivated by the backward-backward splitting method \cite{combettes2011proximal}, similar to previous works \cite{Chun2020ConvolutionalAO}, \cite{chun2020momentum}, we approach the minimization of \eqref{eq:reco_problem} by 
\begin{eqnarray}
\ZZ_j =& \sum_{k=1}^K h_k^\trans \ast \mathcal{S}_{\alpha / \lambda}(h_k \ast \XX_j) \label{eq:nn_steps1} \\
\XX_{j+1} =& \underset{\XX}{\arg \min} \frac{1}{2}\| \Ad\XX - \YY\|_2^2 + \frac{\lambda}{2}\|\XX - \ZZ_j\|_2^2 \label{eq:nn_steps2},
\end{eqnarray}
for $0 \leq j \leq T$, with $\XX_0 := \Ad^\sharp \YY$, where $\Ad^\sharp$ denotes some pseudo-inverse of $\Ad$.  In \eqref{eq:nn_steps1}, $\mathcal{S}_{\alpha/\lambda}$ denotes the soft-thresholding operator with threshold $\alpha/\lambda$ and $h_k^\trans$ denotes the adjoint of $h_k$. Under an orthonormal basis assumption, the sequence defined by \eqref{eq:nn_steps1} and \eqref{eq:nn_steps2} reduces to the backward-backward splitting algorithm for \eqref{eq:reco_problem_no_aux}, known to converge to a minimizer of \eqref{eq:reco_problem} \cite{combettes2011proximal}.
The minimizer of \eqref{eq:nn_steps2} can be obtained by solving a linear system $\Hd\XX = \mathbf{b}_j$ with
\begin{eqnarray}
\Hd =& \Ad^\herm \Ad +   \lambda \, \Id \label{eq:lin_system_H} \\
\mathbf{b}_j =& \Ad^\herm \YY + \lambda \, \ZZ_j \label{eq:lin_system_b},
\end{eqnarray}
where the $\Hd$ does not depend on the filters and is thus computationally favorable.\\
\textbf{Proposed Reconstruction Network}: We propose to train the filters $\{h_k\}_k$ by constructing a network $f_{\Theta}$ which corresponds to a sequence of alternating steps which implement the operations in \eqref{eq:nn_steps1} and \eqref{eq:nn_steps2}. In the network, the filters are treated as trainable parameters, i.e.\ $\Theta = \cup_k \{h_k\}$, and can therefore be learned by back-propagation in a supervised manner on a set of $M$ data-pairs $\mathcal{D} = \{ (\XX_0^i, \XX_{\mathrm{f}}^i)_{i=1}^M \}$, where $\XX_{\mathrm{f}}$ denotes a ground-truth image. Further, we can learn the optimal regularization parameters $\lambda$ and $\alpha$ as well. In order to constrain the regularization parameters to be strictly positive, we apply a Soft-Plus  activation to $\alpha$ and $\lambda$.\\
Because the soft-thresholding operator $\mathcal{S}_{ \alpha / \lambda}$ is not differentiable with respect to its threshold $\alpha / \lambda$, following \cite{Zhang2001}, we smoothly approximate it by 
\begin{equation}
\tilde{\mathcal{S}}_t(z) = z + \frac{1}{2}\bigg( \sqrt{ (z-t)^2 + b} - \sqrt{ (z+t)^2 + b} \bigg)
\end{equation}
to be able to learn the optimal threshold by back-propagation, where $b>0$ is a parameter which we fixed to $b=0.001$. In the network, the complex-valued images are treated as two-channeled real-valued images and the the real and the imaginary parts of the images share the same filters. For the convolutional layers, we employ circular padding. Figure \ref{fig:network} illustrates the proposed network architecture.
\section{Experiments}
\label{sec:experiments}
\begin{figure*}
\centering
\begin{minipage}{\linewidth}
\resizebox{\linewidth}{!}{
\includegraphics[height=2.4cm]{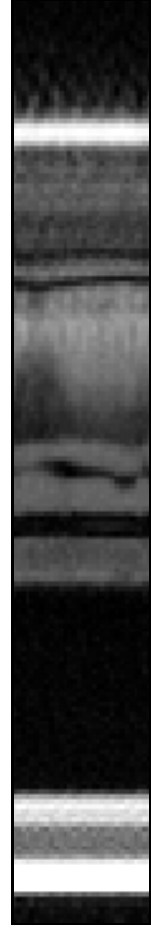}\hspace{-0.1cm}
\begin{overpic}[height=2.4cm,tics=10]{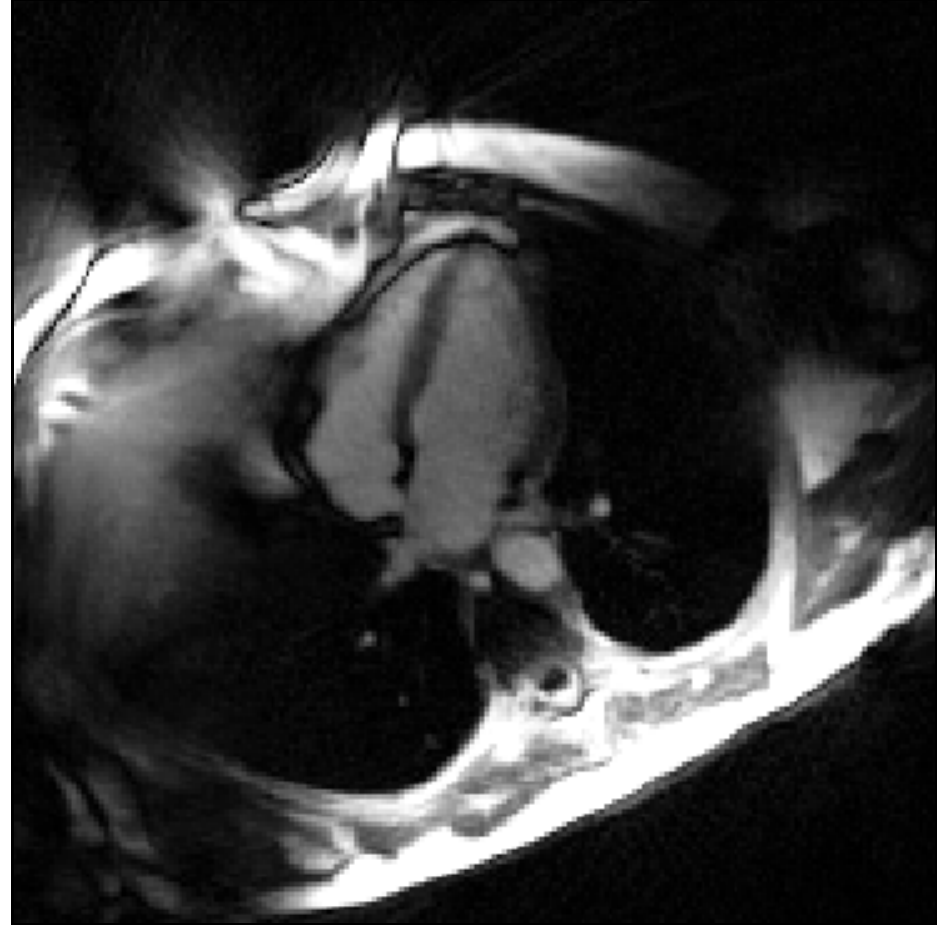}
\put(1,89){ {\textcolor{white}{\scriptsize{\bf{CAOL}}}} }
\end{overpic}
\includegraphics[height=2.4cm]{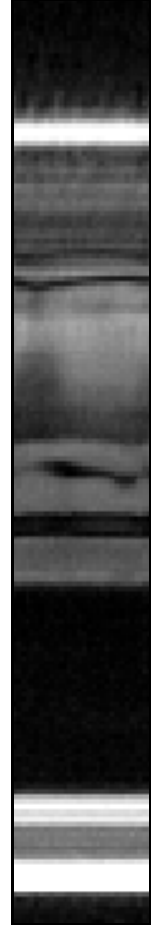}\hspace{-0.1cm}
\begin{overpic}[height=2.4cm,tics=10]{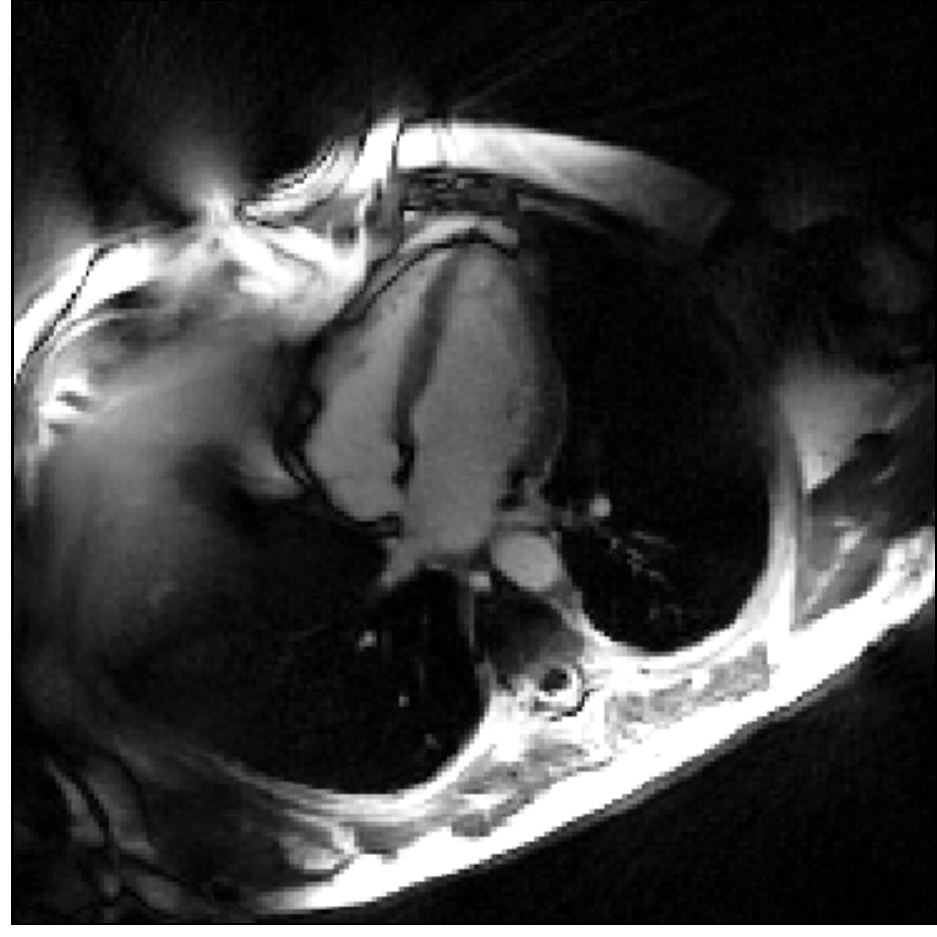}
\put(1,89){ {\textcolor{white}{\scriptsize{\bf{Proposed}}}} }
\end{overpic}
\includegraphics[height=2.4cm]{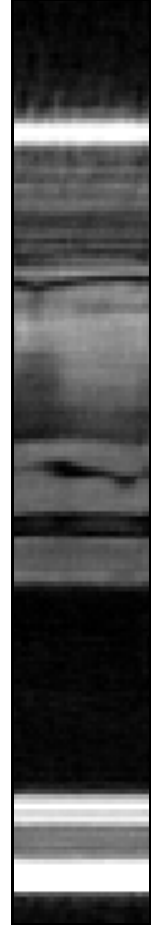}\hspace{-0.1cm}
\begin{overpic}[height=2.4cm,tics=10]{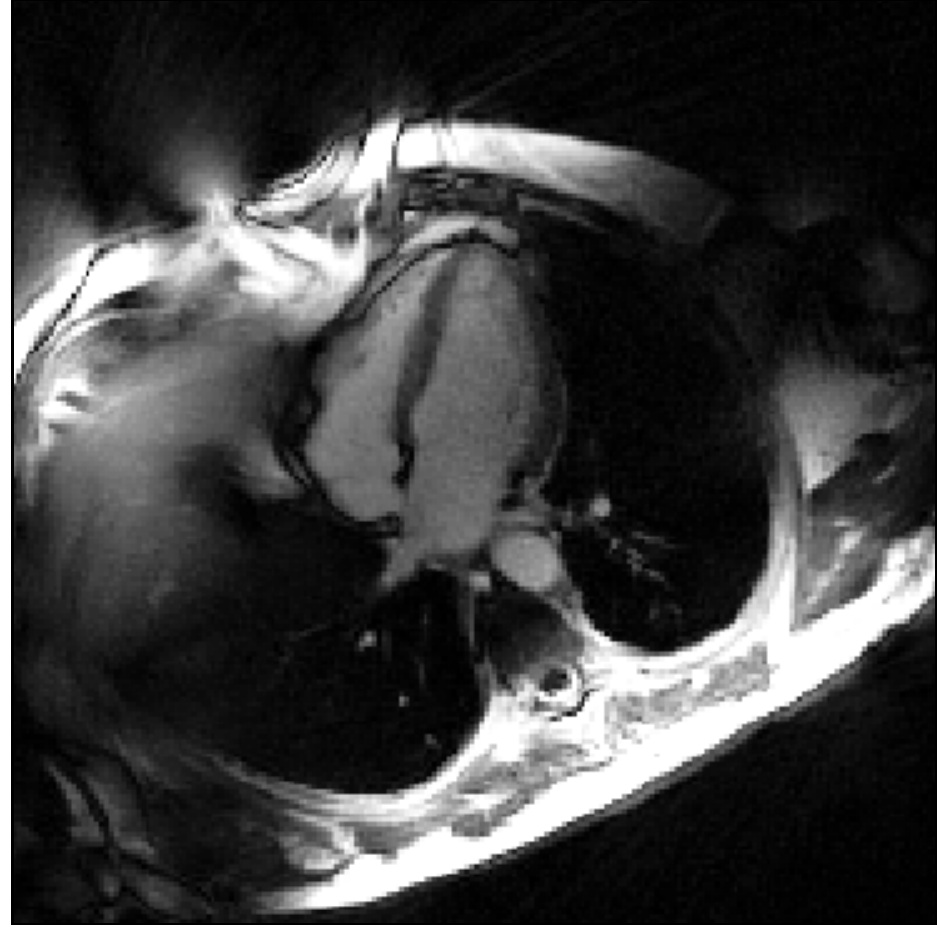}
\put(1,89){ {\textcolor{white}{\scriptsize{\bf{DnCn3D}}}} }
\end{overpic}
\includegraphics[height=2.4cm]{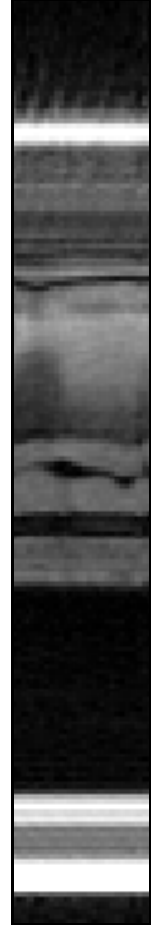}\hspace{-0.1cm}
\begin{overpic}[height=2.4cm,tics=10]{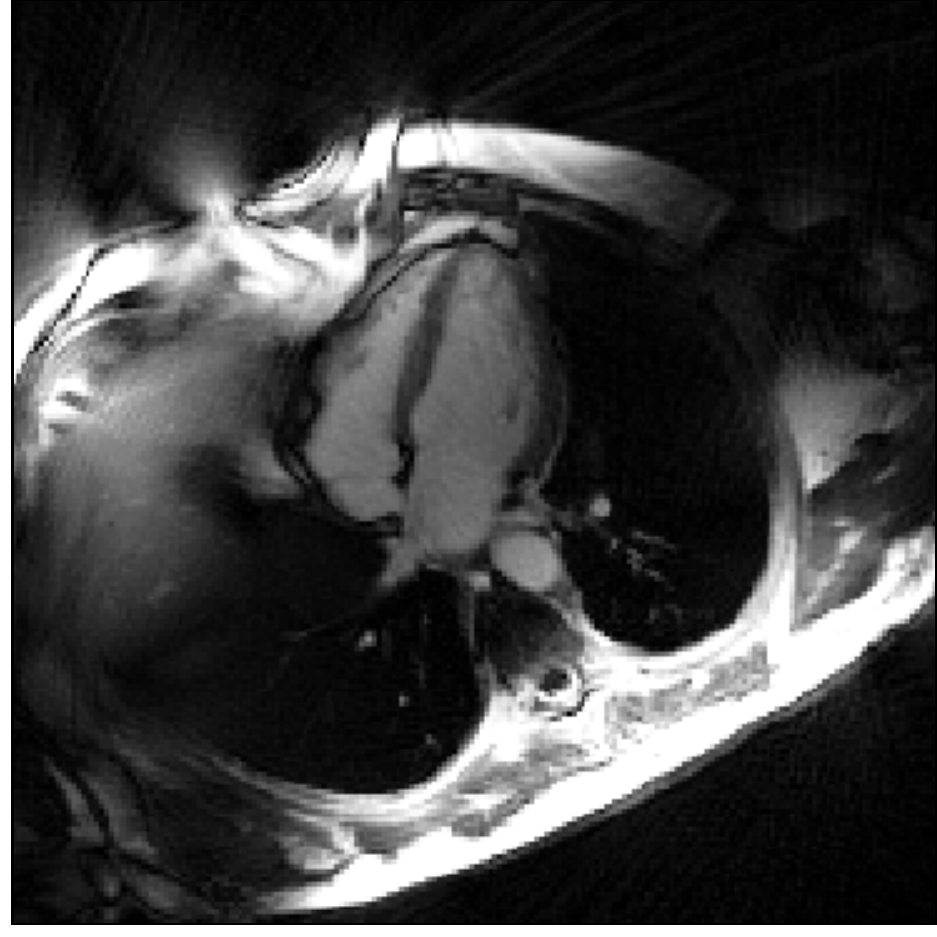}
\put(1,89){ {\textcolor{white}{\scriptsize{\bf{Ground-Truth}}}} }
\end{overpic}
}
\resizebox{\linewidth}{!}{
\includegraphics[height=2.4cm]{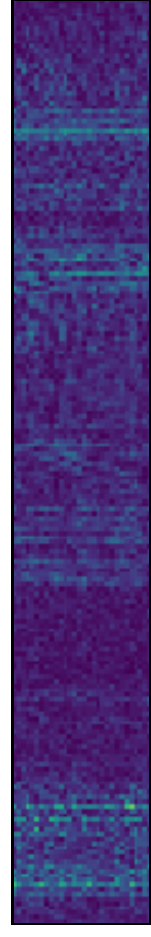}\hspace{-0.1cm}
\begin{overpic}[height=2.4cm,tics=10]{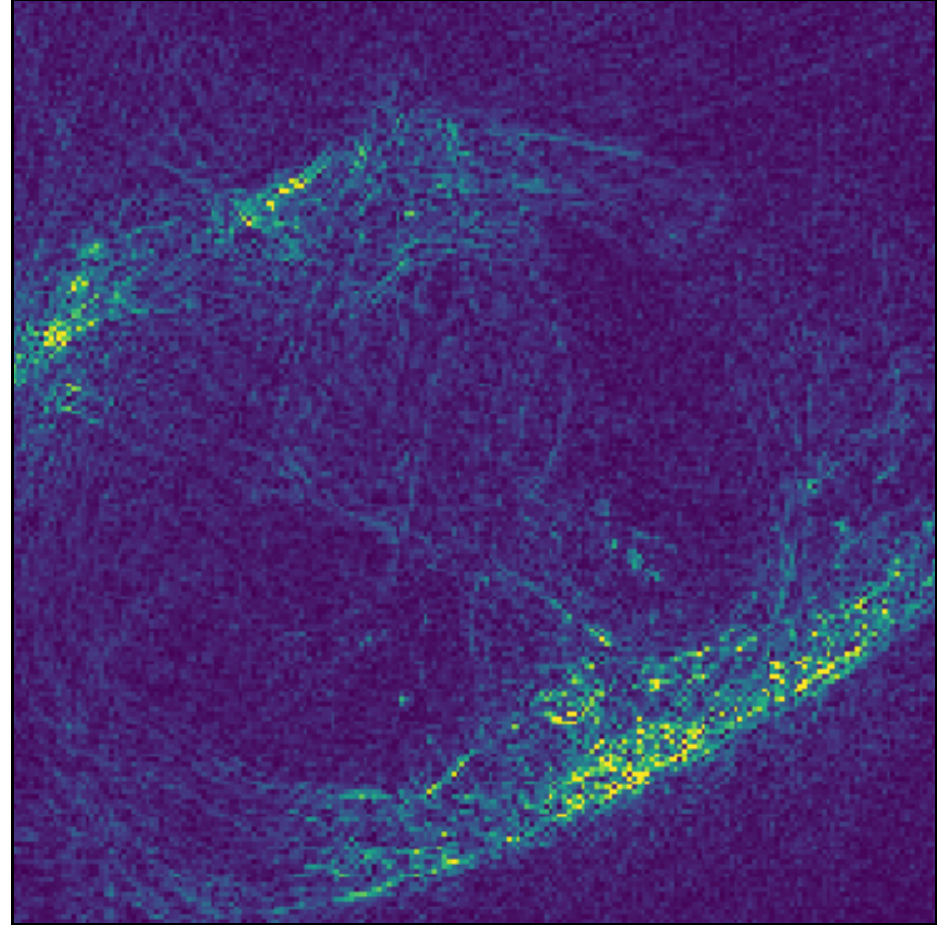}
\end{overpic}
\includegraphics[height=2.4cm]{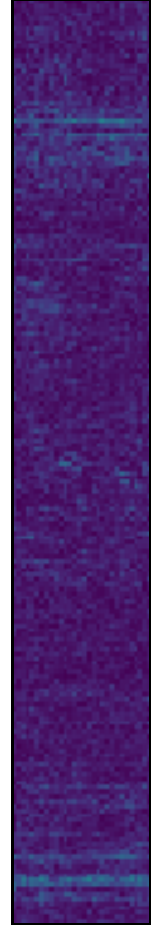}\hspace{-0.1cm}
\begin{overpic}[height=2.4cm,tics=10]{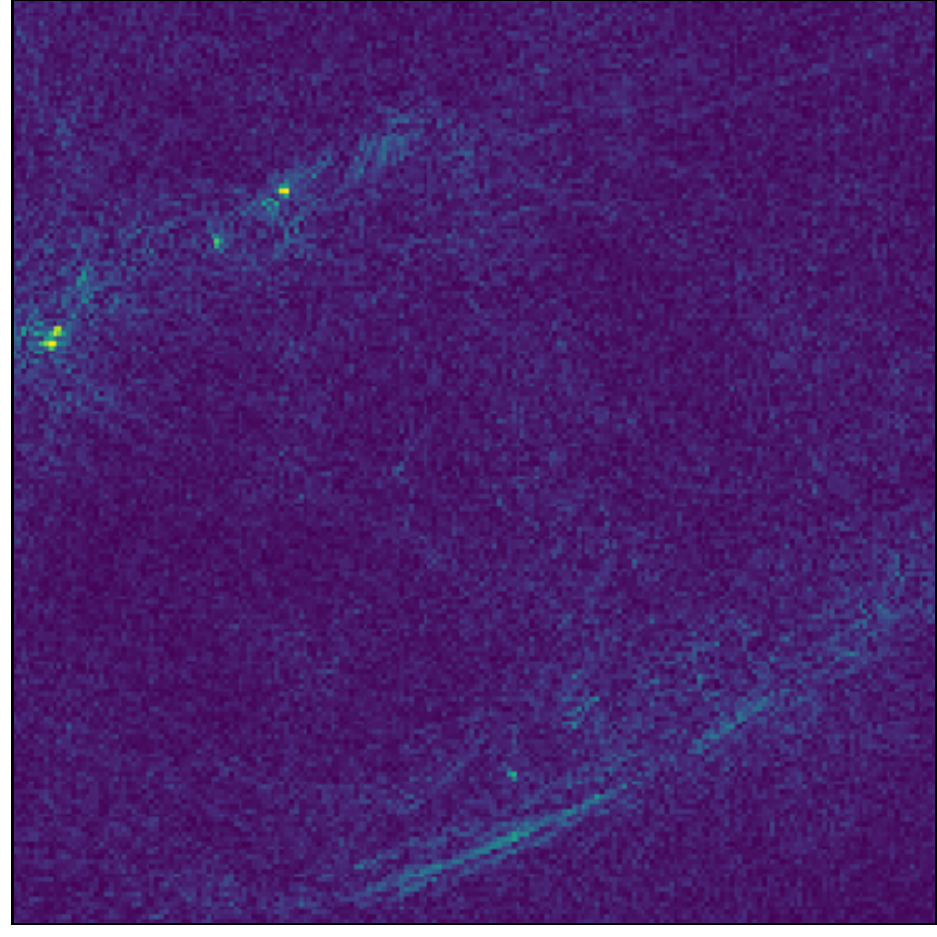}
\end{overpic}
\includegraphics[height=2.4cm]{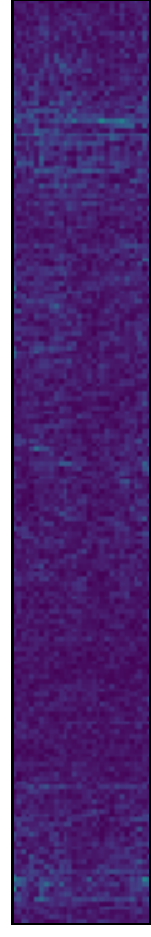}\hspace{-0.1cm}
\begin{overpic}[height=2.4cm,tics=10]{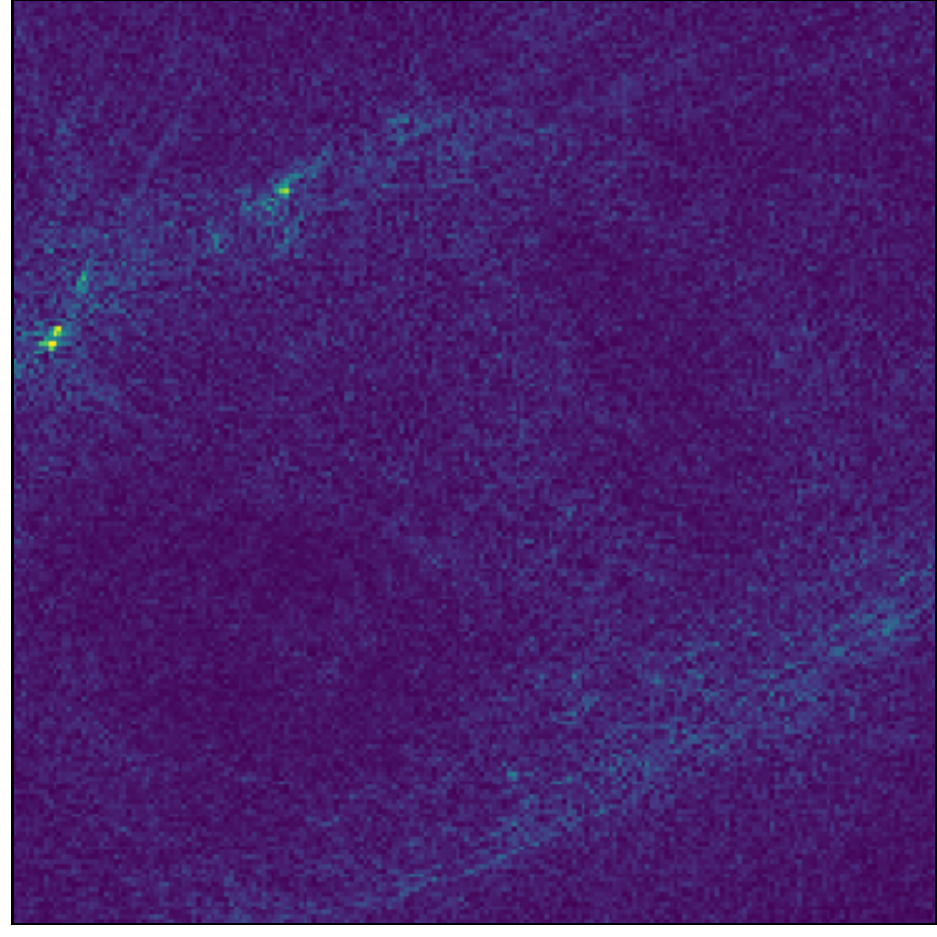}
\end{overpic}
\includegraphics[height=2.4cm]{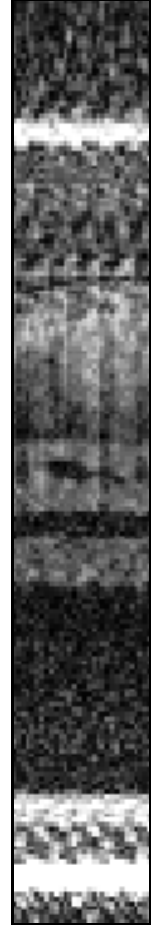}\hspace{-0.1cm}
\begin{overpic}[height=2.4cm,tics=10]{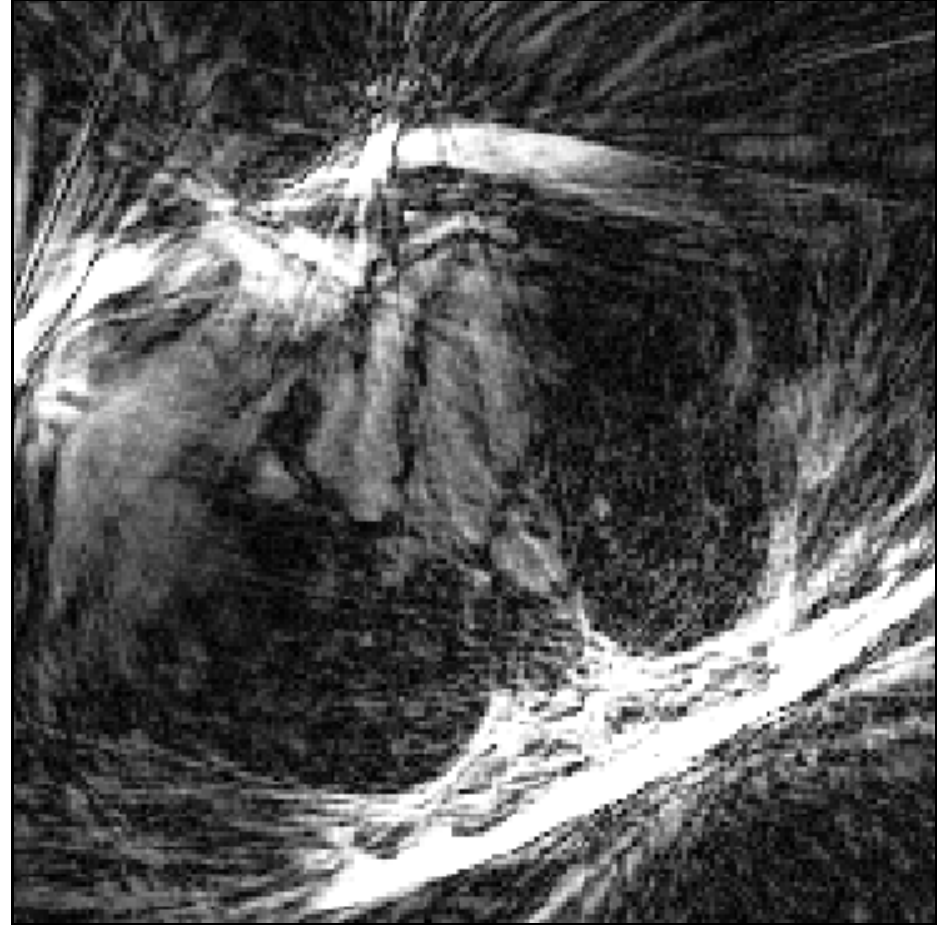}
\put(1,89){ {\textcolor{white}{\scriptsize{\bf{NUFFT}}}} }
\end{overpic}
}
\end{minipage}
\caption{An example of reconstructions and corresponding point-wise error-images of the test set for the proposed reconstruction method using CAOL filters \cite{Chun2020ConvolutionalAO} for $K=16,k_f=3$ and the ones obtained by our proposed end-to-end training-approach for $K=16$ and $k_f=7$ as well as for the deep CNN-cascade DnCn3D \cite{schlemper2017deep}. Although DnCn3D yields a slightly lower point-wise error, our proposed approach shows competitive performance with the advantage that the role of the regularizing kernels is fully interpretable. All results are shown for the best combination of hyper-parameters  (chosen on the validation set) for each respective method.}\label{fig:reco_results}
\end{figure*}
In the following, we tested our proposed method on an accelerated radial cardiac cine MR image reconstruction problem. Similar as in \cite{kofler2021end}, the operator $\Ad$ in  \eqref{eq:inv_problem} is given by
\begin{equation}\label{eq:NUFFTOp}
\Ad:= (\Id_{N_c} \otimes \Ed) \Cd,
\end{equation}
for a complex valued image $\XX = [\XX_1, \ldots, \XX_{N_t}]^\trans \in \mathbb{C}^N$ with $N=N_x \times N_y \times N_t$. The operator $\Id_{N_c}$ denotes an identity operator and $\Cd$ contains the $N_c$ coil-sensitivity maps which are multiplied to the cine MR image, i.e.\ $\Cd =  [\Cd_1,\ldots,\Cd_{N_c}]^\trans$, with $\Cd_j = \mathrm{diag}(\mathbf{c}_j,\mathbf{c}_j,\ldots,\mathbf{c}_j) \in \mathbb{C}^{N \times N}$ and $\mathbf{c}_j \in \mathbb{C}^{N_x \times N_y}$.  The operator $\Ed = \mathrm{diag}(\Ed_1, \ldots, \Ed_{N_t})$ consists of different 2D non-uniform (NUFFT) Fourier-encoding operators $\Ed_t$ which for each point $t \in \{1,\ldots, N_t\}$ sample a 2D image $ \XX_t \in \mathbb{C}^{N_x \times N_y}$ along radial lines in Fourier-space. To accelerate the acquisition process, we only acquire a subset of the $k$-space coefficients which are needed to sample a 2D image $\XX_t$ at Nyquist limit, which we denote by $I\subset J=\{1,\ldots,N_{\mathrm{rad}}\}$. Finally, by $\Au$, we denote  the undersampled 2D radial encoding operator which samples all $k$-space coefficients in the set $I = I_1 \cup \ldots \cup I_{N_t}$ with $I_t \subset J$ for all $t=1,\ldots,N_t$ according to a golden-angle radial pattern \cite{winkelmann2006optimal}. The operator $\Au$ was implemented using \texttt{TorchKBNufft} \cite{Muckley2020}.\\
As often done for non-Cartesian sampling schemes, in the data-consistency term in \eqref{eq:reco_problem_no_aux}, the $k$-space data is multiplied by a diagonal operator $\Wd^{1/2}$ which contains the entries of the density-compensation function and is used to pre-condition the problem. By doing so, the operator $\Au^\sharp$ takes the form $\Au^\sharp:= \Au^\herm \Wd^{1/2}$. Accordingly, in Section \ref{sec:methods}, the operators $\Au^\herm \Au$ in \eqref{eq:lin_system_H} and $\Au^\herm$ in \eqref{eq:lin_system_b} must be replaced by $\Au^\sharp \Au$ and $\Au^\sharp$, respectively.\\
\textbf{Dataset}: We used a set of 15 healthy volunteers and four patients which amounted to 216 cine MR images of shape $320 \times 320 \times 30$. We split the data into 12/3/4 subjects (144/36/36 dynamic images) for training, validation and testing where the test set consisted of the four patients. The initial $k$-space data was retrospectively simulated using an acceleration factor of approximately $R\approx 18$  and $N_c=12$ coil-sensitivity maps and was further corrupted by Gaussian noise with a standard deviation of $\sigma=0.02$.\\
\textbf{Methods of Comparison  and Evaluation}: Since our proposed method is a method for training sparsifying convolutional filters, the first  method of comparison is the one in \cite{Chun2020ConvolutionalAO}, which we denote by CAOL. After having trained the filters with CAOL, we fixed them in our reconstruction network and  only trained the regularization parameters. We also compared our method to a deep cascade of convolutional neural networks \cite{schlemper2017deep}, which we abbreviate by DnCn3D. For DnCn3D, we used $T=4$ and each block has two convolutional layers with $16$ filters, amounting to a total number of 31.232 trainable parameters. Note that the original work in \cite{schlemper2017deep} was presented for a single-coil Cartesian acquisition scheme. For our comparison, we extended the method to be applicable to non-Cartesian multi-coil data-acquisitions by replacing the data-consistency layer in \cite{schlemper2017deep} by a CG module. For CAOL, at test time, the length of the network was increased to $T=24$ as it further decreased the NRMSE . All results were evaluated in terms of PSNR, NRMSE, structural similarity index measure \cite{wang2004image} (SSIM) and universal image quality index \cite{wang2002universal} (UIQ) which were calculated over a central squared ROI of $160 \times 160$ pixels for all cardiac phases.\\
\textbf{Network Training}:  We trained different networks with $K=8,16,24$ for different 3D kernels of shape $k_f\times k_f \times k_f$ for $k_f=3,5,7$ to minimize the squared $L_2$-error between the estimated output and the target-images. Because the application of the NUFFT-operator is computationally expensive and problem \eqref{eq:reco_problem} is separable with respect to the time points, for our method, we reduced the number of cardiac phases to $N_t=8$ during training. We set $T=4$ and used $n_{\mathrm{CG}}=4$ iterations to solve \eqref{eq:nn_steps2}. All methods were trained using the ADAM optimizer with an initial learning rate of $10^{-4}$. Our network was trained for 75 epochs ($\approx 9$ hours), while DnCn3D was trained for 500 epochs ($\approx 4$ days).
\begin{figure}[t]
\centering
\includegraphics[width=0.49\linewidth]{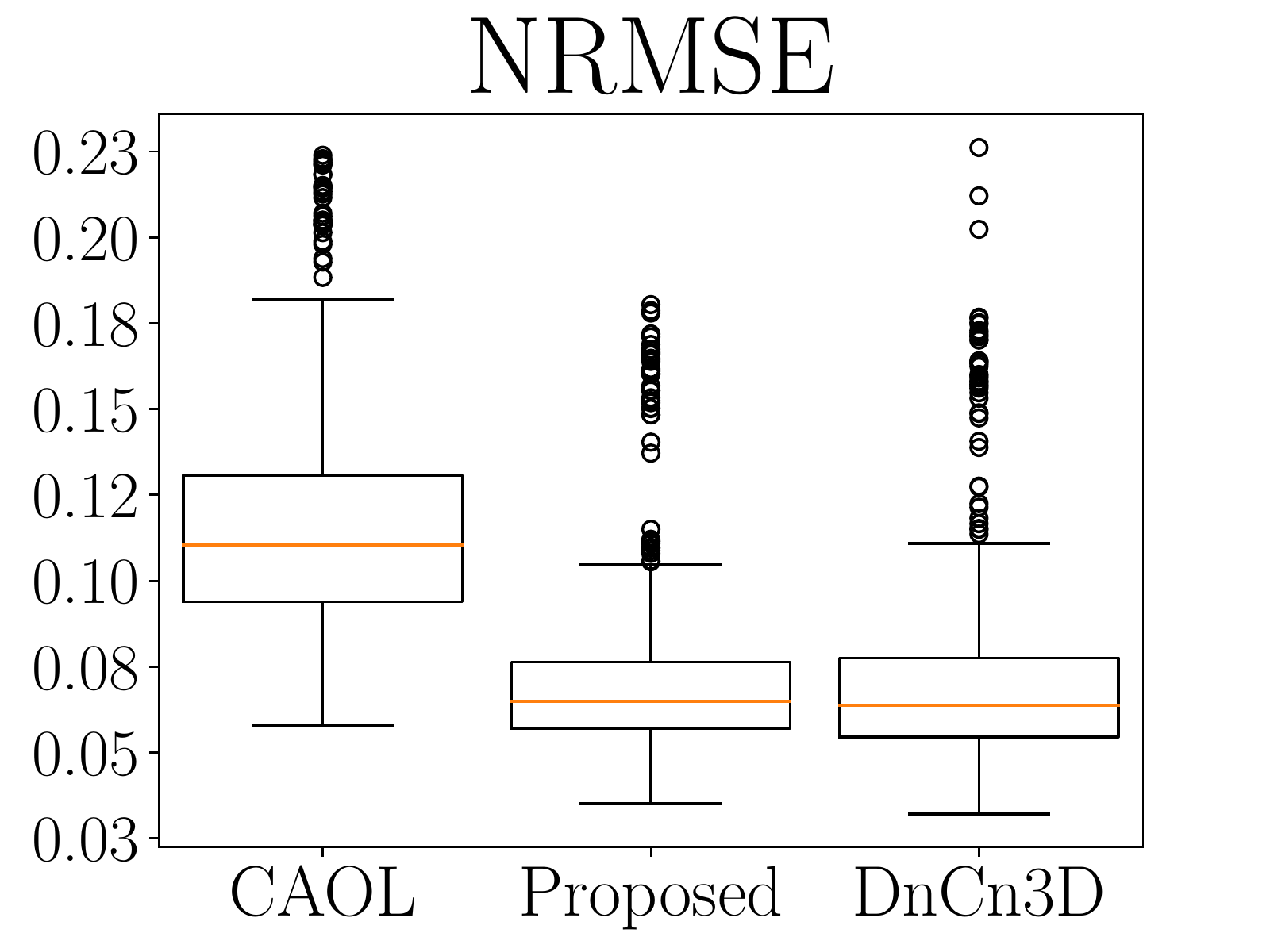}
\includegraphics[width=0.49\linewidth]{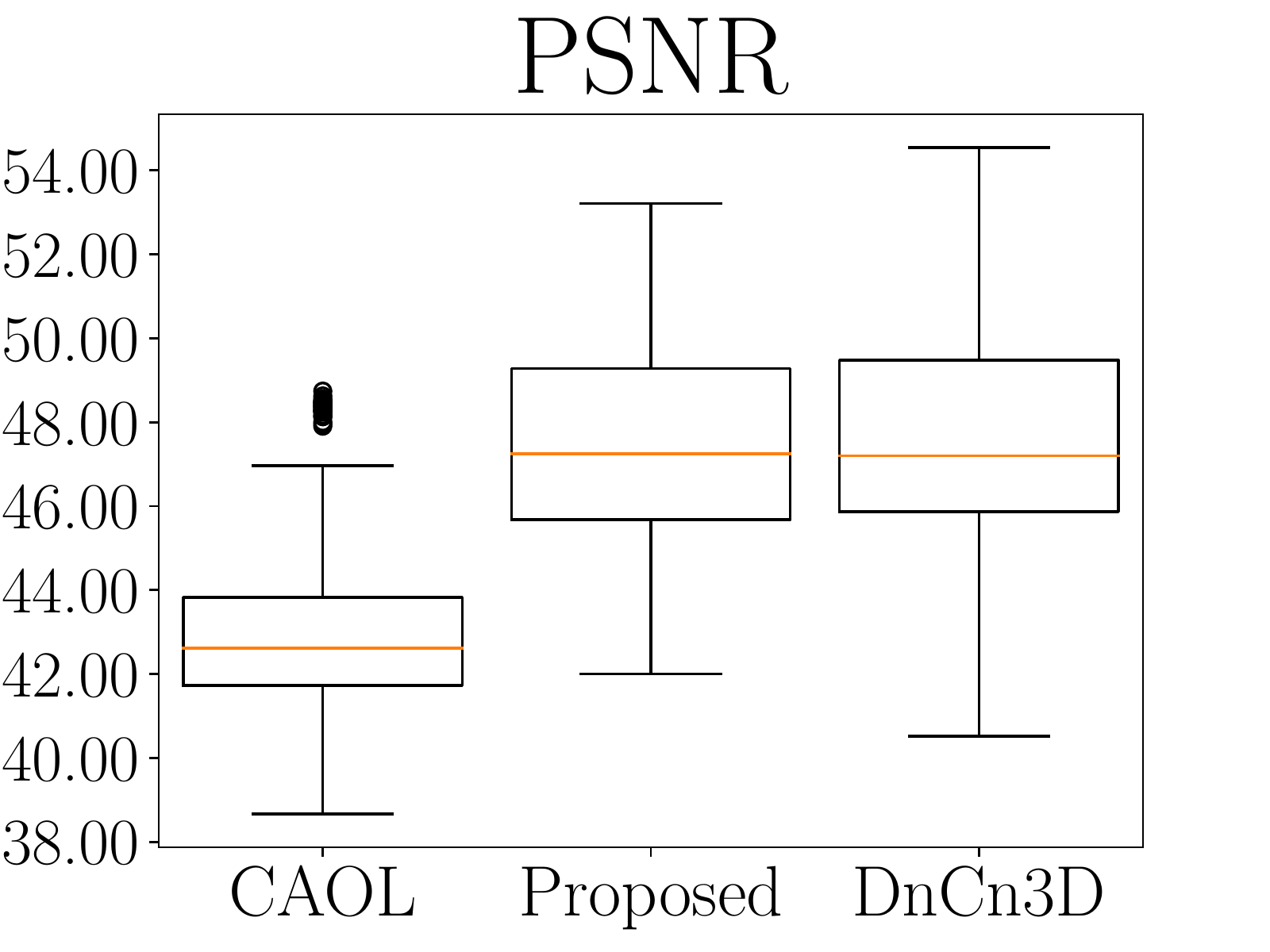}\\
\includegraphics[width=0.49\linewidth]{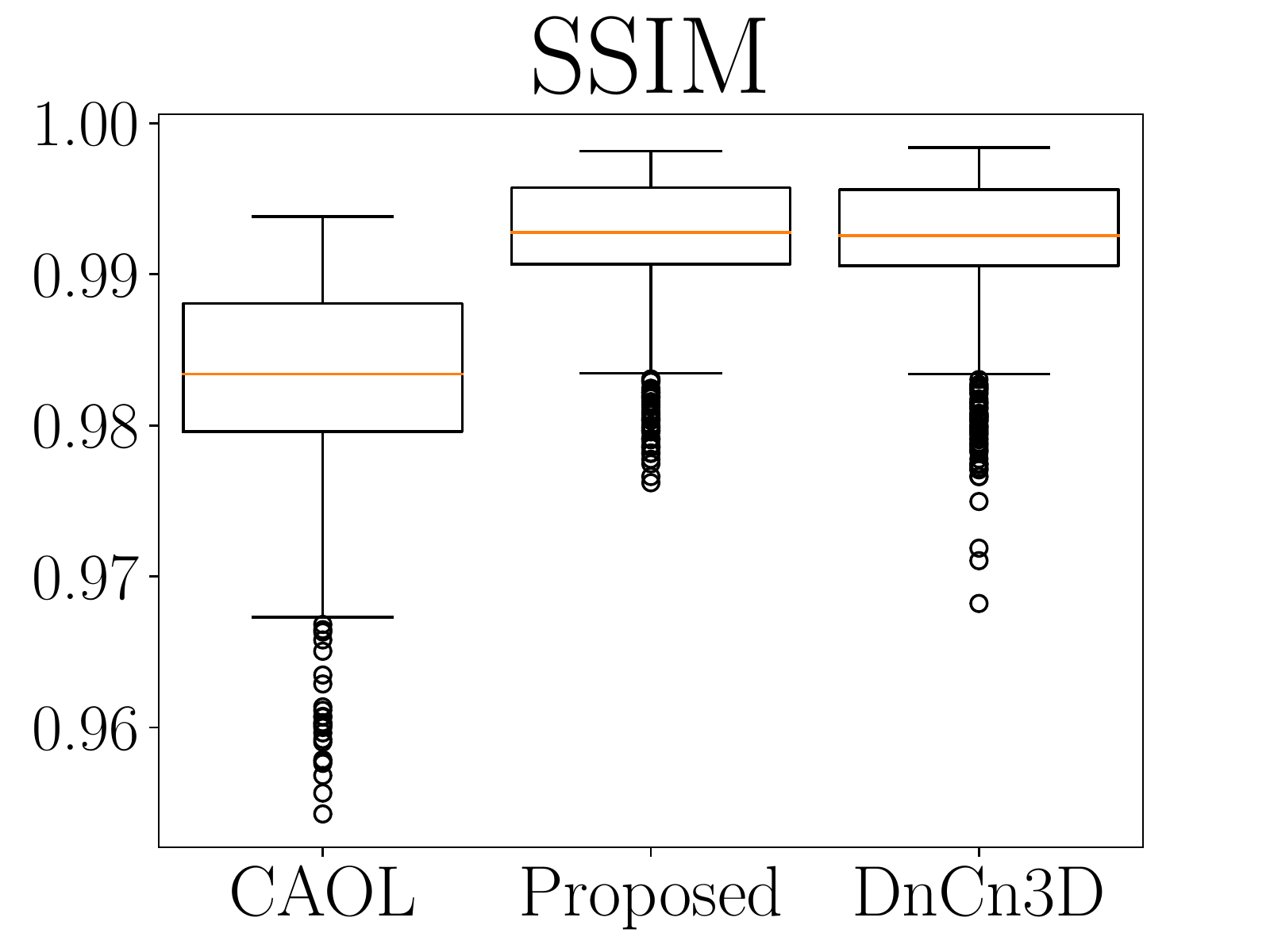}
\includegraphics[width=0.49\linewidth]{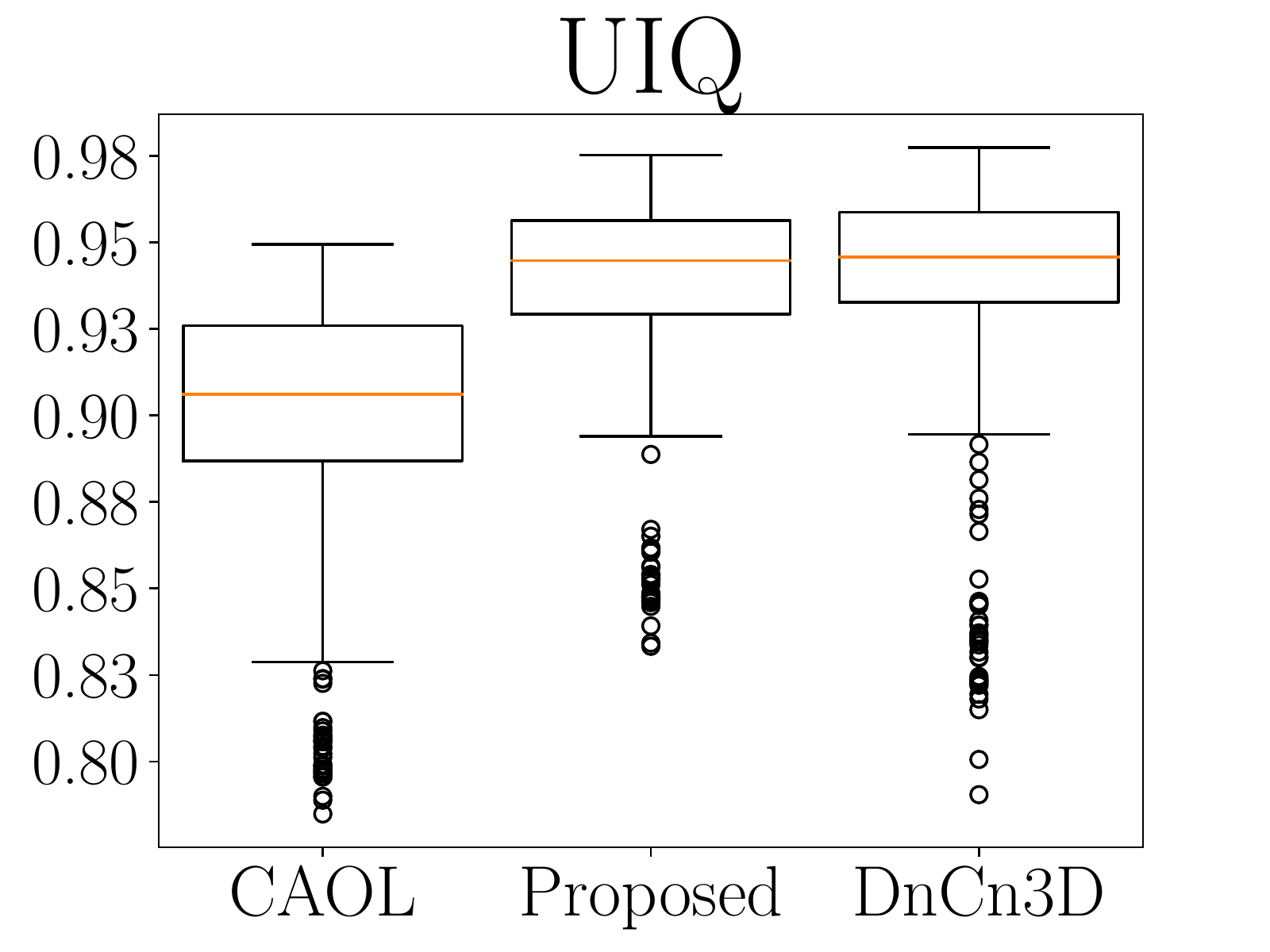}
\caption{Box-plots of the quantitative results obtained for CAOL \cite{Chun2020ConvolutionalAO} for $K=16,k_f=3$, our proposed method for $K=16,k_f=7$ and DnCn3D \cite{schlemper2017deep}. Our proposed method yields similar results as \cite{schlemper2017deep} while only having $110<K/2\cdot k_f^3 + 2<4.118$ trainable parameters (i.e.\ the filters and the regularization parameters $\alpha$ and $\lambda$) compared to 31.233 for DnCn3D. }\label{fig:box_plots}
\end{figure}
\begin{figure}[!ht]
\centering
\includegraphics[width=0.7\linewidth]{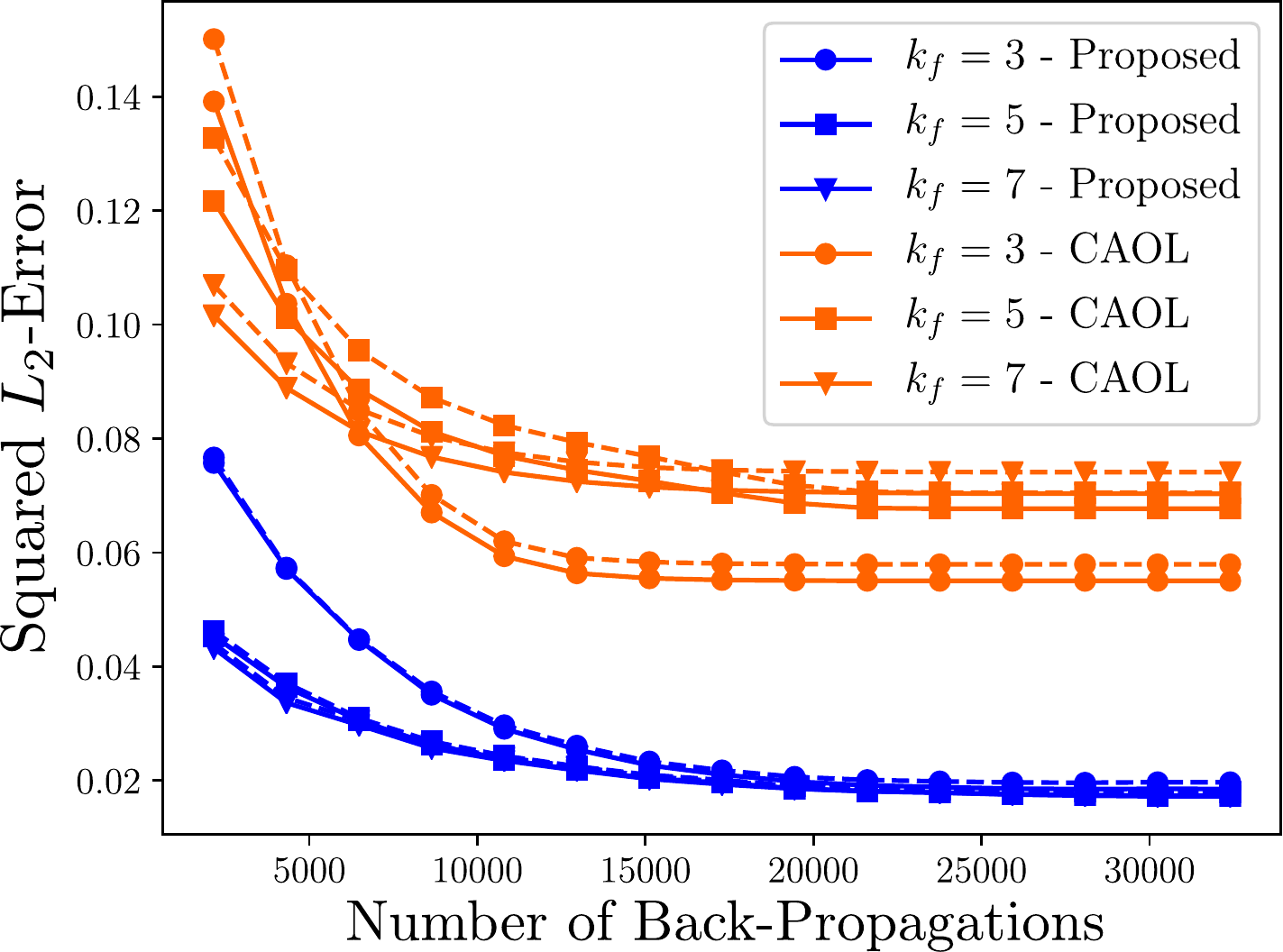}
\caption{Training- and validation-error (solid/dashed) during the optimization of the convolutional filters with our proposed reconstruction network for $K=16$ and different $k_f$ (only shown for $K=16$ for presentation purposes). For CAOL, only $\lambda$ and $\alpha$ were trained. }\label{fig:training_curves}
\end{figure}
\section{Results and Discussion}
\label{sec:results}
In Figure \ref{fig:reco_results}, we see an example of a   reconstruction for our method using the CAOL-filters with $K=16$ and $k_f=3$ and the ones obtained by end-to-end training, which yield a visibly smaller point-wise error and better preserve image details. These results are also supported in terms of the reported quantitative measures, as can be seen in the box-plots in Figure \ref{fig:box_plots}. Our proposed method and DnCn3D clearly surpass CAOL. Further, the proposed reconstruction method yields comparable results to DnCn3D and further seems to be slightly more stable, as can be seen from the outliers in the box-plots. This can  most probably be attributed to the fact that it contains significantly fewer trainable parameters.  Note that the results for CAOL and our proposed method are shown for the best configuration of $K$ and $k_f$ based on the validation set. This can be seen from Figure \ref{fig:training_curves} which shows the training and validation errors for our network and for CAOL for $K=16$. Increasing the filter size $k_f$ slightly reduces the achievable validation error for our method.    Interestingly, we found that CAOL performs better with smaller kernel-sizes. Although this might seem somewhat counter-intuitive, this aspect shows that choosing the optimal hyper-parameters for decoupled methods is challenging. In contrast, using iterative networks to train the convolutional filters, larger filter-sizes tend to lead to smaller reconstruction errors and the filters are optimally adjusted to be used with the employed reconstruction algorithm regardless of the chosen hyper-parameters.
\section{Conclusion}
\label{sec:conclusion}
In this work, we have shown that end-to-end trained iterative neural networks can be used to learn classical sparsity-based regularization methods in a task-driven and physics-informed manner. The obtained sparsifying transforms are better tailored to the employed reconstruction algorithm compared to the ones obtained by the corresponding decoupled method. Further, we have evaluated our method on a realistic large-scale dynamic cardiac MR problem and found that the proposed method yields  results which are on par with  the ones obtained by a state-of-the-art method employing a deep cascade of neural networks. In addition, in our method, the exact role of the regularizer is fully explainable and allows for a theoretical analysis of the reconstruction algorithm which we leave for future work.
Although we have presented the approach for a dynamic non-Cartesian MR image reconstruction example, we point out that the method may be applicable to other imaging modalities as well.

\section{Acknowledgments}
\label{sec:acknowledgments}
No funding was received for conducting this study. The authors have no relevant financial or non-financial interests to disclose.
\section{Compliance with Ethical Standards}
\label{sec:compl}
All subjects gave written informed consent before participation, in accordance with the ethical committee of the responsible institution.
\bibliographystyle{IEEEbib}
\bibliography{strings,refs}
\end{document}